\documentclass[a4paper]{jpconf}
\usepackage{graphicx}
\usepackage{amsmath}
\usepackage{amssymb}

\begin{document}
\title{Cosmic neutrinos at IceCube: $\theta_{13}$, $\delta$ and initial
flavor composition}

\author{Arman Esmaili}

\address{Department of Physics, Sharif University of Technology, P.O.Box 11365-8639, Tehran,
IRAN}
\address{School of Physics, Institute for Research in Fundamental Sciences (IPM), P.O.Box 19395-5531, Tehran, IRAN}

\ead{arman@mail.ipm.ir}

\begin{abstract}
We discuss the prospect of extracting the values of the mixing
parameters $\delta$ and $\theta_{13}$ through the detection of
cosmic neutrinos in the planned and forthcoming neutrino
telescopes. We take the ratio of the $\mu$-track to shower-like
events, $R$, as the realistic quantity that can be measured in the
neutrino telescopes. We take into account several sources of
uncertainties that enter the analysis. We then examine to what
extent the deviation of the initial flavor composition from
$w_e:w_\mu:w_\tau=1:2:0$ can be tested.

\end{abstract}



The neutrino mixing parameters can be extracted from cosmic
neutrinos data according to the following argument
\cite{parameters}: assume that the flavor ratio of neutrinos at
the source is $w_e:w_\mu:w_\tau$; after that neutrinos travel the
large distances between the astrophysical sources and the Earth
the oscillatory terms in the flavor transition probabilities
average out such that the flavor ratio at the detector will become
\begin{equation} \sum_\alpha w_\alpha P_{\alpha e}:\sum_\alpha
w_\alpha P_{\alpha \mu}:\sum_\alpha w_\alpha P_{\alpha \tau},
\end{equation} where
\begin{equation} P_{\alpha \beta}\equiv P(\nu_\alpha \to
\nu_\beta)=P(\bar{\nu}_\alpha \to \bar{\nu}_\beta)=\sum_i
|U_{\alpha i}|^2|U_{\beta i}|^2, \end{equation} and $U_{\alpha i}$
are the elements of the neutrino mixing matrix. In a wide range of
models the flavor ratios at the source are predicted to be
$w_e:w_\mu:w_\tau=1:2:0$. Thus, by measuring the flavor ratio at
Earth, one can derive the absolute values of the mixing matrix
elements which in principle yield information on the yet-unknown
neutrino parameters $\theta_{13}$ and $\delta$.


IceCube \cite{icecubesensitivity} and its Mediterranean
counterparts ( such as KM3NET \cite{KM3NET}) can basically
distinguish only two types of events: 1) shower-like events; 2)
$\mu$-track events. It is possible to derive information on the
flavor composition of neutrinos by studying the ratio $R=$number
of $\mu$-track events/number of shower-like events \cite{Beacom}.
In the analysis in this paper we consider neutrinos with energies
100~GeV~$<E_\nu<$~100~TeV, where the upper and lower limits come
from the absorption of neutrinos in Earth and the energy threshold
of detection in neutrino telescopes, respectively.

Two sources contribute to the $\mu$-track events: (i) Charged
Current (CC) interaction of $\nu_\mu$ or $\bar{\nu}_\mu$ producing
$\mu$ or $\bar{\mu}$; (ii) CC interaction of $\nu_\tau$ and
$\bar{\nu}_\tau$ producing $\tau$ or $\bar{\tau}$ and the
subsequent decay of $\tau$ and $\bar{\tau}$ into $\mu$ and
$\bar{\mu}$. In the literature, the contribution of $\nu_\tau$
(via $\nu_\tau \to \tau \to \mu$) to $\mu$-track events has been
overlooked but to study the effect of $\theta_{13}$, one should
take into account such sub-dominant effects (see the Appendix of
\cite{Esmaili:2009dz} for details). Three types of events appear
as shower: i) the Neutral Current (NC) interactions of all kinds
of neutrinos; ii) the CC interactions of $\nu_e$ and
$\bar{\nu}_e$; iii) the CC interactions of $\nu_\tau$
($\bar{\nu_\tau}$) and the subsequent hadronic decay of $\tau$
($\bar{\tau}$). Details of the event rate calculation for each
case in both $\mu$-track and shower-like events can be found in
\cite{Esmaili:2009dz}.

\begin{figure}[h!]

  \begin{minipage}{0.5\textwidth} \includegraphics[bb=200 70 570 590,width=6cm,height=7cm,clip=true,angle=-90]{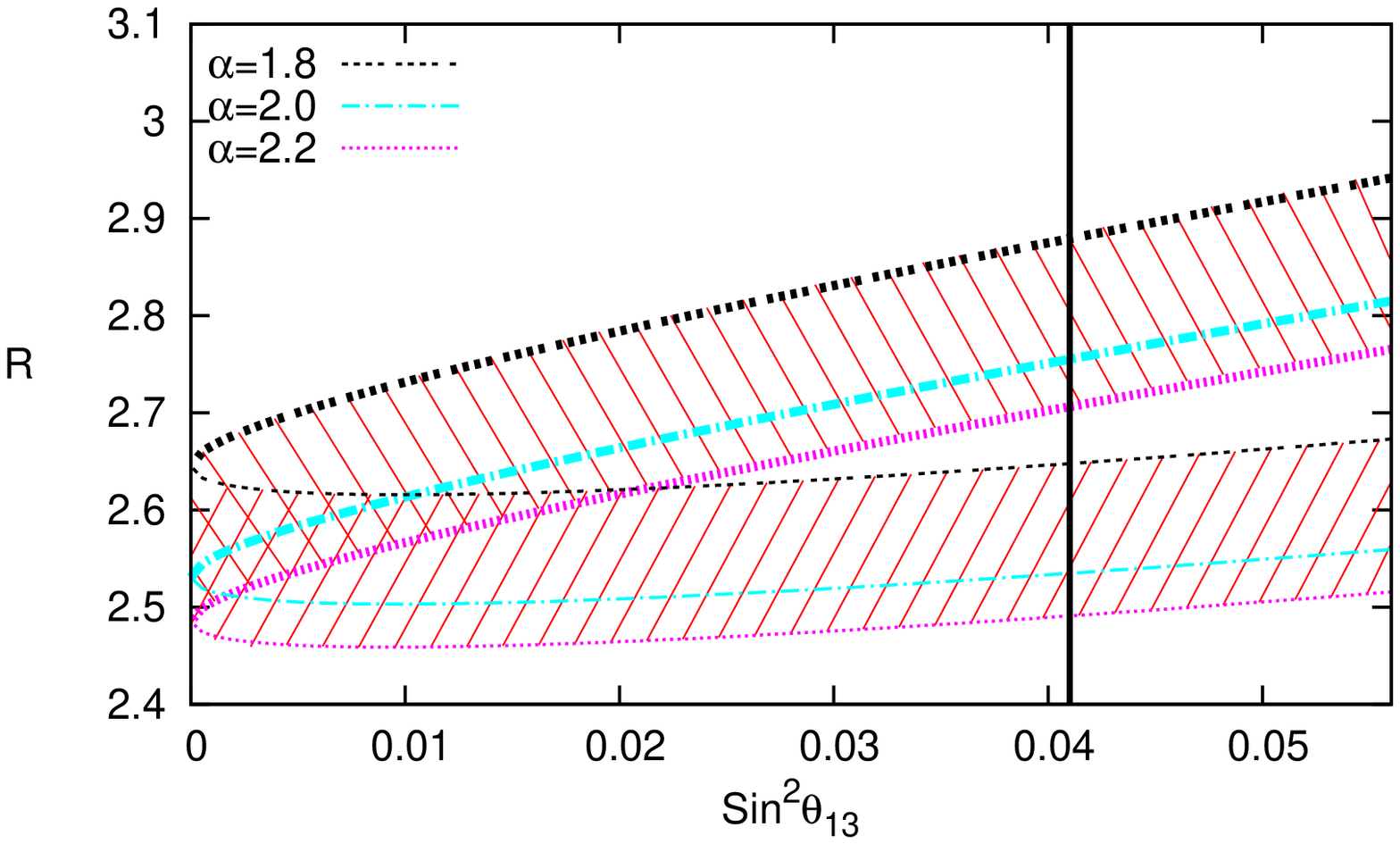}
  \end{minipage} \hspace{1.2cm}
  \begin{minipage}{0.5\textwidth}
  \includegraphics[bb=200 70 580
  600,width=6cm,height=7cm,clip=true,angle=-90]{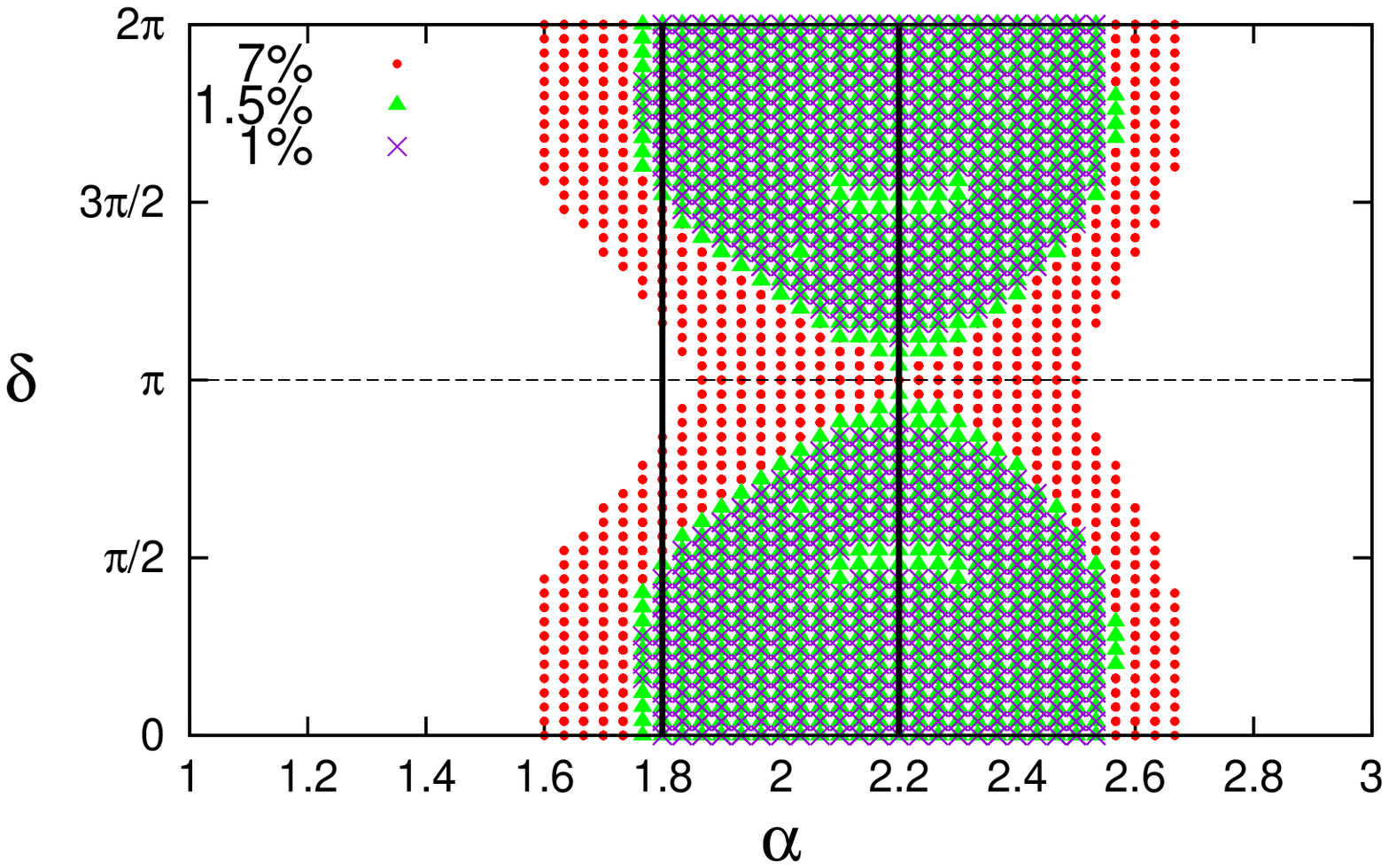}
  \end{minipage}
  \centerline{\hspace{0.8cm}(a)\hspace{9.0cm}(b)}
  \caption{{\small (a) The dependence of $R$ on $\sin^2\theta_{13}$
  for different values of the spectral index, $\alpha$. The thicker lines correspond to
  $\delta=\pi$ and the thinner ones correspond to $\delta=0$.
  In drawing this figure we have set $\mathcal{N}_{\bar{\nu}_e}/\mathcal{N}_{\nu_e}=0.5$ and
  $(\mathcal{N}_{\bar{\nu}_\mu}+\mathcal{N}_{\nu_\mu})/(\mathcal{N}_{\bar{\nu}_e}+\mathcal{N}_{\nu_e})=2$.
  The input for $\theta_{12}$ and $\theta_{23}$ are set equal to the best
  fit in \cite{schwetz}.
  The vertical line at 0.041 shows the present bound at 3$\sigma$ \cite{schwetz}
  (b) Points in the $(\alpha,\delta)$ space
  consistent with $R=2.53 \pm \Delta R$. True values of the
  $(\alpha,\delta)$ pair are $(2,\pi/2)$. Points displayed by dots,
  triangles and crosses respectively correspond to
  $ \Delta R/\bar{R}=7\%$, $ \Delta
   R/\bar{R}=1.5\%$ and $ \Delta
  R/\bar{R}=1\%$. To draw this figure we have varied
  $\sin^2\theta_{13} \in (0.028,0.032)$,
  $\sin^2\theta_{12} \in (0.30,0.34)$, $\sin^2\theta_{23} \in (0.47,0.53)$
  and $\mathcal{N}_{\bar{\nu}_e}/\mathcal{N}_{{\nu}_e}\in(0,1)$.} }
  \label{fig1,2}
\end{figure}

Several input parameters enter the calculation of $R$ and their
uncertainties induce imprecision in the calculated value of $R$.
We consider the following uncertainties in the calculation of $R$:
i) for the energy spectrum of neutrinos we assume a power-law
spectrum $dF_{\nu_\beta}/dE_{\nu_\beta}=\mathcal{N}_{\nu_\beta}
E^{-\alpha}_{\nu_\beta}$. Neutrino production through the Fermi
acceleration mechanism of particles in the source predict the
value of the spectral index equal to $\alpha=2$, but taking into
account non-linear effects in the acceleration mechanism results
in spectral index values $\alpha\in(1,3)$. It is shown in
\cite{Beacom} by assuming the flux
$E_\nu^2dF_\nu/dE_\nu=0.25$~GeV~cm$^{-2}$~sr$^{-1}$~yr$^{-1}$,
after one year of data-taking $\alpha$ can be determined with 10\%
uncertainty. For the normalization factor
$\mathcal{N}_{\nu_\beta}$ it can be shown that:
$\mathcal{N}_{\bar{\nu}_\mu}=\mathcal{N}_{\nu_\mu}$ and
$\mathcal{N}_{\bar{\nu}_e}/\mathcal{N}_{\nu_e}\in(0,1)$.

\begin{figure}[h!]
  \begin{center}
 \centerline{\includegraphics[bb=260 50 570
 500,keepaspectratio=true,clip=true,angle=-90,scale=0.43]{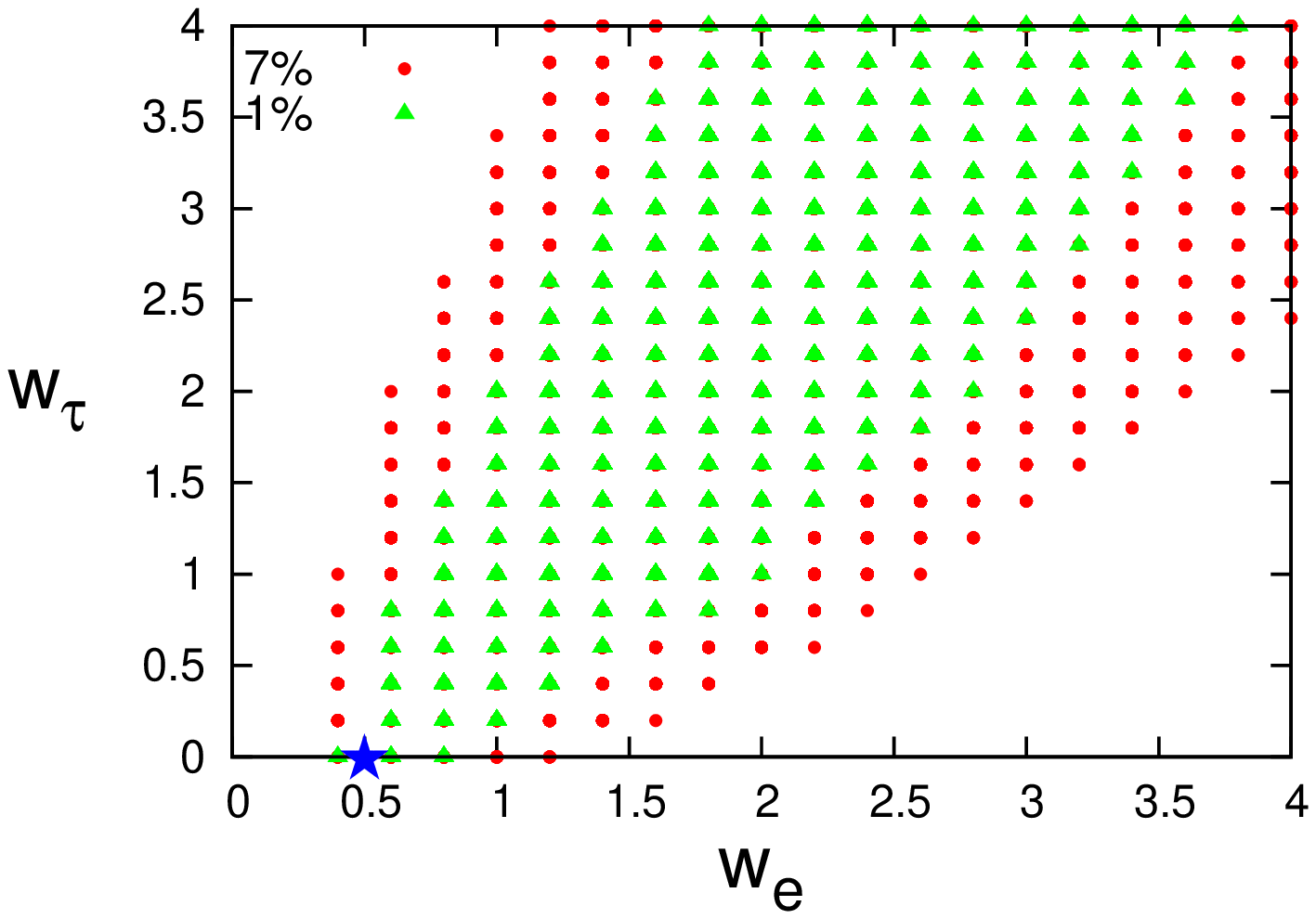}\includegraphics[bb=260 40 570
 500,keepaspectratio=true,clip=true,angle=-90,scale=0.43]{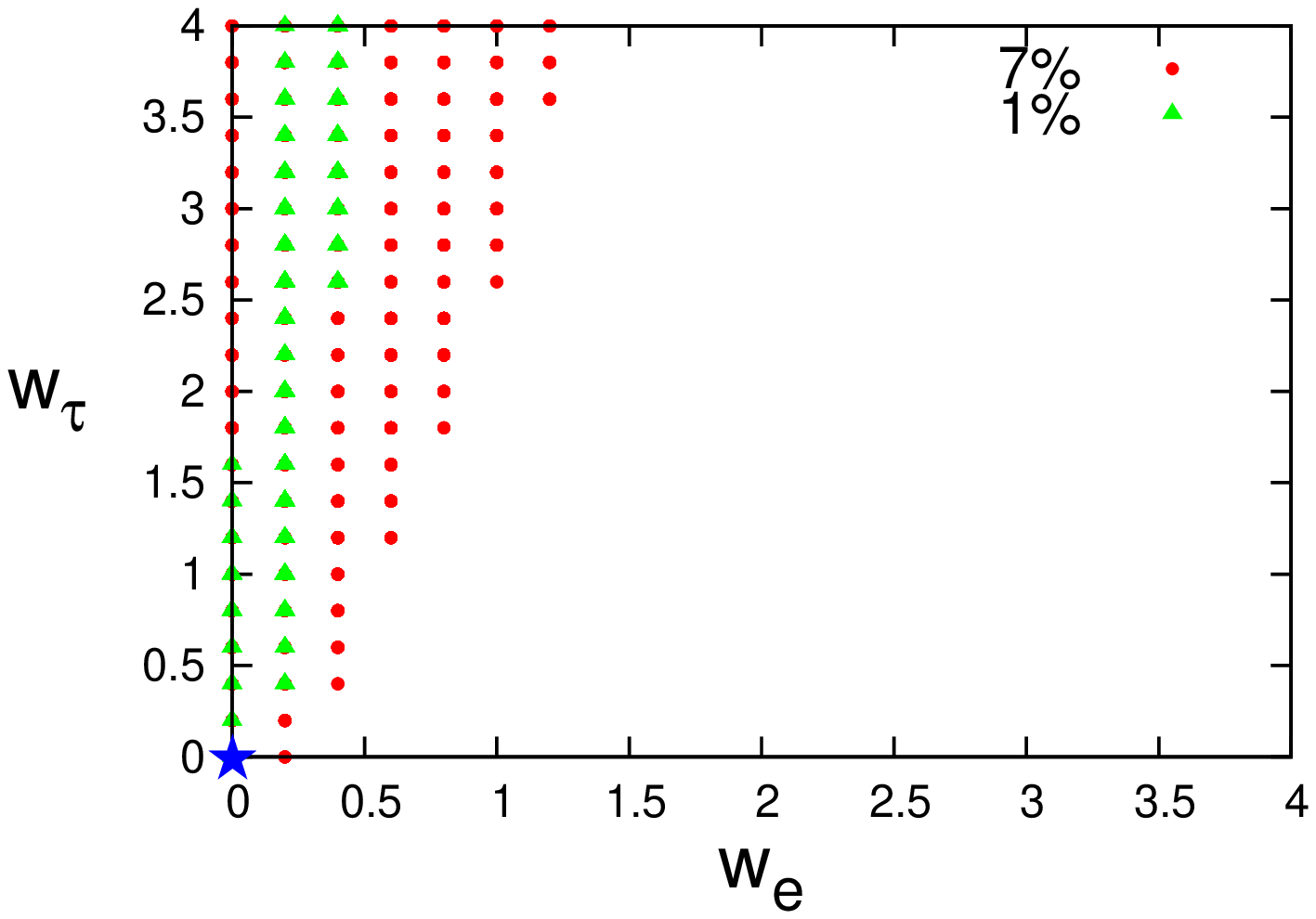}}
 \centerline{\vspace{-0.45cm}}
 \centerline{\hspace{0.8cm}(a)\hspace{6.8cm}(b)}
 \centerline{\vspace{-1.6cm}}
 \end{center}
 \caption{{\small Points in the $(w_e,w_\tau)$ plane consistent
 with $R=\bar{R} \pm \Delta R$. The ratios are normalized such that
 $w_\mu=1$. The true values of $(w_e,w_\tau)$ are denoted by
 $\bigstar$. Points displayed by dots and
 triangles respectively correspond to $ \Delta R/\bar{R}=7\%$ and $ \Delta
 R/\bar{R}=1\%$. In drawing this figure we have varied
 $\sin^2\theta_{13} \in (0,0.003)$, $\delta \in (0,2\pi)$,
 $\alpha\in (1.8,2.2)$,
 and $\mathcal{N}_{\bar{\nu}_e}/\mathcal{N}_{{\nu}_e}\in(0,1)$.
 Drawing Fig.~(a), we have taken $\bar{R}=2.53$ which corresponds
 to the standard picture with $w_e=1/2$ and $w_\tau=0$. In case
 Fig.~(b), we have set $\bar{R}=3.2$ which corresponds to the
 stopped muon scenario with $w_e=w_\tau=0$.}} \label{ab120}
\end{figure}

Fig.~(\ref{fig1,2}-a) shows $R$ versus $\sin^2 \theta_{13}$ for
$\cos \delta=\pm 1$ and various values of $\alpha$. As seen from
the figure when $\delta=0$, the sensitivity of $R$ to $s_{13}^2$
is very mild and less than 2~\%. That is while for $\cos
\delta=-1$, the sensitivity to $s_{13}^2$ is about 10~\%. As seen
from the figure, even for $\cos \delta=-1$, the sensitivity to
$s_{13}^2$ can be obscured by the 10~\% uncertainty in $\alpha$.
However, for $s_{13}^2>0.02$, the bands between $\alpha=2.2$ and
$1.8$ for $\cos \delta=1$ and $\cos \delta=-1$ have no overlap.
This means that for $s_{13}^2>0.02$, 10~\% precision in $\alpha$
is enough to distinguish $\cos \delta=1$ from $\cos \delta=-1$.

Fig.~(\ref{fig1,2}-b) addresses the question that whether it will
be possible to extract the value of $\delta$. Drawing the plot, we
have assumed that $\bar{R}$ will be found to have a typical value
of 2.53 with an uncertainty of $\Delta R/\bar{R}$. This value of
$\bar{R}$ can be obtained by taking maximal CP-violation
$(\delta=\pi/2)$, $\sin^2\theta_{13}=0.03$,
$w_e:w_\mu:w_\tau=1:2:0$, $\alpha=2$ and
$\mathcal{N}_{\bar{\nu}_e}/\mathcal{N}_{{\nu}_e}=0.5\ $. We have
looked for solutions in the $\delta-\alpha$ plane for which
$R=2.53(1\pm\Delta R/\bar{R})$, varying the rest of the relevant
parameters in the ranges indicated in the caption of
Fig.~(\ref{fig1,2}-b). The regions covered with dots, little
triangles and crosses respectively correspond to 7\%, 1.5\% and
1\% precision in the measurement of $R$. As seen from the figure,
with $\Delta R/\bar{R}=7\%$, $\delta$ cannot be constrained. In
fact, any point between the vertical lines can be a solution. The
figure shows that reducing $\Delta R/\bar{R}$ to 1\% (but keeping
the rest of the uncertainties as before), some parts of the
solutions can be excluded. In particular, the region around
$\delta=\pi$ will not be a solution anymore. Notice that along
with $\delta =\pi/2$, $\delta=0$ is also a solution. This means
that despite maximal CP-violation, the CP-violation cannot still
be established.

Taking into account the relevant uncertainties in the input
parameters, we look for values of $w_e:w_\mu:w_\tau$ that are
consistent with $R=\bar{R} \pm \Delta R$. To perform this
analysis, we take $\theta_{13}=0$. In Fig.~(\ref{ab120}), we
consider two possibilities: (i) the standard case with
$w_e:w_\mu:w_\tau=0.5:1:0$ leading to $\bar{R}=2.53$ (see
Fig.~\ref{ab120}-a); (ii) the case of stopped muons with
$w_e:w_\mu:w_\tau=0:1:0$ yielding $\bar{R}=3.20$ (see
Fig.~\ref{ab120}-b). From these figures we observe that with a
precision of $\Delta R/\bar{R}=7\%$, these two scenarios can be
easily discriminated. These two can also be discriminated from the
scenario in which the neutrino production mechanism is $n\to p e
\bar{\nu}_e$ ({\it i.e.,} $w_e:w_\mu:w_\tau=1:0:0$). When we
restrict the analysis to $w_\tau=0$ ({\it i.e.,} the case without
exotic neutrino properties) from these figures we observe that the
measurement of $R$ stringently constrains $w_e:w_\tau$ which in
turn sheds light  on the production mechanism. However, once the
assumption of $w_\tau$ is relaxed, a wide range of
$w_e:w_\mu:w_\tau$ can be a solution. For example, the exotic case
of $w_e:w_\mu:w_\tau=0:0:1$ leads to the same value of $\bar{R}$
as the stopped muon scenario.

\ack I am grateful Y.~Farzan for useful discussions and comments.
I would like to thank ``Bonyad-e Melli-e Nokhbegan'' for partial
financial support.

\end{document}